\documentclass{emulateapj}

\slugcomment{Draft Version August 31, 2006: Submitted to ApJ}

\usepackage{times}
\usepackage{bm}

\newcommand{\bolB}{{\bm  B}}
\newcommand{\bolJ}{{\bm  J}}

\shorttitle{Non-axisymmetric Magnetic Tower Jets}
\shortauthors{Nakamura et al.}

\begin{document}

\title{Stability Properties of Magnetic Tower Jets}

\author{Masanori Nakamura\altaffilmark{1}, Hui Li\altaffilmark{1}, 
and Shengtai Li\altaffilmark{2}}
  \altaffiltext{1}{Theoretical Astrophysics, MS B227, Los Alamos
  National Laboratory, NM 87545; nakamura@lanl.gov}
  \altaffiltext{2}{Mathematical Modeling and Analysis, MS B284, 
  Los Alamos National Laboratory, NM 87545}

\begin{abstract}
Stability  properties  of ``magnetic  tower''  jets  propagating in  the
gravitationally stratified  background have been  examined by performing
three-dimensional       magnetohydrodynamic       simulations.       The
current-carrying,  Poynting  flux-dominated  magnetic tower  jet,  which
possesses  a highly  wound helical  magnetic  field, is  subject to  the
current-driven instability  (CDI). We find that,  under general physical
conditions  including  small  perturbations  in the  initial  background
profiles,   the   propagating    magnetic   tower   jets   develop   the
non-axisymmetric, $m=1$  kink mode of the  CDI.  The kink  mode grows on
the local Alfv\'en  crossing time scale. In addition,  two types of kink
modes  appear in  the  system.   At the  central  region where  external
thermal pressure confinement  is strong, only the internal  kink mode is
excited and  will grow.  A large  distance away from  the central region
where the external thermal pressure  becomes low, the external kink mode
is observed.  As  a result, the exterior of magnetic  tower jets will be
deformed  into  a  large-scale   wiggled  structure.   We  also  discuss
extensively  the different  physical  processes that  contribute to  the
overall stability  properties of the magnetic  tower jets. Specifically,
when the jet propagates in  an initially unperturbed background, we find
that they  can survive the kink  mode beyond the point  predicted by the
well-known Kruskal-Shafranov (K-S) criterion.  The stabilization in this
case  comes mainly  from  the dynamical  relaxation  of magnetic  twists
during the propagation of magnetic towers; the magnetic pitch is reduced
and  the corresponding  K-S critical  wavelength becomes  longer  as the
tower jet  proceeds.  Furthermore, we show that  the pressure-driven and
Kelvin-Helmholtz instabilities do not  occur in the magnetic tower jets.
This  strongly suggests  that  the CDI  is  the primary  reason for  the
wiggling structures in jets.
\end{abstract}


\section{INTRODUCTION}  
\label{sec:INT} Magnetohydrodynamic  (MHD) mechanisms are  often invoked
to  explain the  launching, acceleration  and collimation  of  jets from
Young Stellar  Objects, X-ray  binaries, Active Galactic  Nuclei (AGNs),
Microquasars,  and  Quasars   \citep[see,  {\it  e.g.},][and  references
therein]{DLM01}.   Strongly magnetized jets,  particularly those  with a
strong toroidal field encircling the collimated flow, are often referred
to as ``current-carrying'' or ``Poynting flux-dominated'' (PFD) jets.  A
large  current flowing  parallel  to  the jet  flow  is responsible  for
generating a  strong, tightly wound helical magnetic  field.  The global
picture of a  current-carrying jet with a closed  current system linking
the magnetosphere of the central engine and the hot spots was introduced
by \citet[]{B78,  B06} and  applied to AGN  double radio  sources.  This
closed  current  system  includes  a  pair  of  current  circuits,  each
containing both  a forward electric  current path (the jet  flow itself,
with  its  toroidal magnetic  field,  toward  the  lobe), and  a  return
electric current path (along some path back to the AGN core).

Recent high resolution observations,  in X-ray, optical and radio bands,
show that many  AGN and quasar jets display  wiggles/kinks, which may be
associated with  a helical motion of  the underlying bulk  flow, in wide
spatial ranges from sub-parsec  to kpc scales \citep[{\it e.g.},][]{R89,
H92, B97, F99, LZ01, M01, EO02, S06}.  Many ideas have been proposed for
these   helical    distortions:   MHD   instabilities    such   as   the
Kelvin-Helmholtz  instability   (KHI),  the  current-driven  instability
(CDI), and  the pressure-driven instability (PDI), or  the precession of
the  jet ejection  axis due  to  the existence  of a  binary black  hole
\citep[]{B80}, or an encounter with another galactic core.

It  is  well known  that  a cylindrical  plasma  column  with a  helical
magnetic configuration  is subject to  the $m=0$ (sausage) mode  of PDI,
the $m=1$  (kink) mode of CDI,  and the other higher  order modes, where
$m$ is  the azimuthal ($\phi$) mode number  \citep[see, {\it e.g.},][for
details]{K66,  BA78, F82}.  CDI  is driven  by parallel  (axial) current
flows and can  exist even in a zero pressure,  force-free plasma. On the
other  hand, PDI is  powered by  perpendicular current  flows and  it is
often known as  the interchange (internal) modes.  KHI  is considered to
be very important  at the shearing boundary between  the flowing jet and
the external  medium, particularly  in the kinetic  flux-dominated (KFD)
jets.   In principle,  CDI  and  KHI have  both  external (surface)  and
internal (body) modes.  PFD jets should be especially susceptible to CDI
rather  than PDI and  KHI.  This  is because  the strong  axial electric
current is  present inside  the PFD jets,  which is responsible  for the
highly wound  helical field configurations, and the  Lorentz force plays
an  important  role in  the  jet  dynamics.   The investigation  of  the
destructive influence of CDI on PFD  jet flows is an important avenue of
research in the study of astrophysical jets.

Much  more  attention  on  studies  of stabilities  (disruption  of  the
interior/exterior of astrophysical jets) have  given to KHI than to CDI.
This is because jets were  believed to be super-Alfv\'enic or super-fast
magnetosonic  KFD flows  ({\it  i.e.}, kinetic  energy  was expected  to
greatly  exceed the magnetic  energy).  However,  it is  not necessarily
always the case; a  super-Alfv\'enic or super-fast magnetosonic PFD flow
can exist  in both nonrelativistic \citep[]{KS97,  V00} and relativistic
\citep[]{VK01} MHD outflow solutions.  Analytic  studies on the CDI as a
possible  explanation for  jet disruption  have been  performed  as well
\citep[]{E93,  S97,  B98,  L99}.   Linear  analyses  of  nonrelativistic
force-free  jets,   which  are   thermally  confined  by   the  external
nonmagnetized medium,  were performed  in super-fast \citep  []{AC92} or
trans-fast  \citep[]{A96}  magnetosonic regimes.   They  found that  the
growth rate of CDI is always  substantially smaller that those of KHI in
their  specific  treatment.   Only  for quite  small  fast  magnetosonic
numbers,  $M_{\rm  F} \lesssim  0.5$,  both  become comparable.   Linear
analyses   of   the   relativistic   force-free   jets,   in   case   of
constant-longitudinal field  $B_{z}$ along the jet  axis, were performed
by \citet{IP94, IP96}.  They concluded that such a jet is stable against
CDI, while \citet[]{L99} analyzed  the relativistic force-free jets with
$B_{z}$ decreasing outwards and found that they are unstable.

Nonlinear development  of the CDI for cold  super-fast magnetosonic jets
has  been   studied  by  \citet{L00},  based  on   the  linear  analysis
\citep[]{A00}.  It  was found that the current  density is redistributed
within the inner  part of the jet radius on  a characteristic time scale
of the order of the Alfv\'en crossing time in the jet frame.  Nothing in
their numerical  results indicated a  possible disruption of the  jet by
the CDI  sausage ($m=0$) or kink ($m=1$)  modes \citep[see also][]{LF00,
F00}.  In  general, these linear/nonlinear  considerations conclude that
``current-driven  instabilities are  unlikely  to be  dangerous for  the
integrity of astrophysical jets'' \citep[]{A96} and this has been widely
accepted  in theoretical/observational  communities.   In the  nonlinear
regime,  an interplay between  KHI and  CDI can  occur in  the helically
magnetized flow even with slightly super-fast magnetosonic speed $M_{\rm
F} \gtrsim 1$ \citep[]{BK02}.  This nonlinear interaction can contribute
to jet  survival, and  the large-scale magnetic  deformations associated
with  the CDI  mode  development can  effectively  saturate KHI  surface
vortices and prevent jet disruption.

By contrast, the first attempt  to investigate the nonlinear behavior of
visible distortion in  KFD jets along the large-scale  magnetic field by
the CDI  kink mode was performed  by \citet[]{T93} to  model Herbig Haro
objects.  In  a similar study,  \citet[]{N01} examined the  formation of
wiggled structures by the $m=1$ mode  in PFD jets to apply for AGN jets.
Their numerical result is  applicable for the systematic distribution of
the observed  rotation measure in 3C 449  \citep[]{F99}, associated with
the  jet deformation \citep[]{K04}.   The rotating  PFD/KFD jets  can be
potentially  stabilized  against  the  $m=1$ mode  beyond  the  original
Kruskal-Shafranov criterion, however, sudden destabilization of rotating
jets  due to  the $m=1$  mode can  occur via  the angular  momentum loss
triggered by MHD shock waves \citep[]{NM04}.

The purpose of the present  paper is to discuss the stability properties
of  ``magnetic tower'' jets,  which have  not been  studied in  the past
work.  Of  particular interest here is  the nonlinear growth  of the CDI
kink  mode.   The   CDI  kink  mode  in  magnetic   tower  jets  may  be
distinguished from that in the  classical MHD jets along the large-scale
magnetic field  \citep[]{T93, N01, NM04}, while it  should be comparable
directly to ``non-disrupting'' models  \citep[]{A96, A00, L00}, in which
the  MHD  jets  are  assumed  to  be in  pressure  equilibrium  with  an
unmagnetized ambient medium.

This  is the  third of  our series  of papers  to examine  the nonlinear
magnetic tower jets.  The  first of our series, \citet[][hereafter Paper
I]{L06},  described the  basic  assumptions and  the  approaches in  the
numerical modeling of  magnetic tower jets.  The evolution  of the tower
jets in a constant  density/pressure background has been examined there.
The second of our series, \citet[][hereafter Paper II]{N06} investigated
the  global structure  of  magnetic tower  jets  in the  gravitationally
stratified  atmosphere (a  more realistic  astrophysical  situation), in
terms of the MHD waves  structure, the radial force equilibrium, and the
collimation.   In  the  present  paper,  the third  of  our  series,  we
investigate the stability properties of magnetic tower jets.  We examine
our numerical  results in two  cases; one without  initial perturbations
and  another  with  initial  perturbations.   The  former  exhibits  the
quasi-axisymmetric  evolution without any  growth of  MHD instabilities,
while  the latter  shows non-axisymmetric  evolution by  the  CDI modes.
This paper is  organized as follows.  In \S 2,  we outline our numerical
methods.  In  \S 3, we  describe our numerical results.   Discussion and
conclusion are given in \S 4 and \S 5.

\section{NUMERICAL METHODS AND MODEL ASSUMPTIONS}
We  solve the  nonlinear system  of time-dependent  ideal  MHD equations
numerically  in a  3-D Cartesian  coordinate system  $(x,\,y,\,z)$.  The
basic  numerical  treatments (including  the  MHD  numerical scheme)  is
essentially the  same as that in Paper  I and II.  We  assume an initial
hydrostatic  equilibrium  in   the  gravitationally  stratified  medium,
adopting an iso-thermal King model \citep[]{K62}.  The magnetic flux and
the  mass are  steadily  injected in  a  central small  volume during  a
certain  time  period.   Since  the  injected magnetic  fields  are  not
force-free, they will  evolve as a ``magnetic tower''  and interact with
the ambient medium.

\subsection{Numerical Set Up}
In this paper, we present mainly  two different cases: one is called the
``unperturbed case'' in the following discussion, which is a run without
any  initial  perturbation  to  the background  initial  profiles.  This
simulation is  taken from the  run presented in  Paper II. The  other is
called the ``perturbed case'',  where a finite amplitude perturbation (a
few percent of the background sound speed) is given to the velocities of
the background gas.

The  total computational  domain is  taken to  be  $|x|,\,|y|,\,|z| \leq
16$. The  numbers of  grid points in  the simulations reported  here are
$240^3$ in the unperturbed case and $320^3$ in the perturbed case, where
the  grid  points  are assigned  uniformly  in  the  $x$, $y$,  and  $z$
directions. We normalize physical  quantities with the unit length scale
$R_{0}$, the  unit density $\rho_{0}$,  the sound speed $C_{\rm  s0}$ as
the typical velocity  field in the system, and  other quantities derived
from  their  combinations,  {\it  e.g.},  the typical  time  $t_{0}$  as
$R_0/C_{\rm s0}$,  etc. Normalizing factors  are $R_0 = 5$  kpc, $C_{\rm
s0} = 4.6 \times 10^7$ cm/s,  $t_{0} = 1.0 \times 10^7$ yrs, $\rho_{0} =
5.0  \times 10^{-27}$  g/cm$^3$, the  unit  pressure $p_0$  as $  \rho_0
C_{\rm s0}^2  = 1.4 \times  10^{-11}$ dyn/cm$^2$, and the  unit magnetic
field  $B_{0}$ as $\sqrt{4  \pi \rho_0  C_{\rm s0}^2}=17.1$  $\mu$G.  We
take  $\rho_0$ and  $p_{0}$  as  the initial  quantities  at the  origin
$(x,\,y,\,z)=(0,\,0,\,0)$ and  the $\rho$ and  $p$ at the origin  in the
simulations are set to unity.  The initial sound speed in the simulation
is  constant,  $C_{\rm  s}=\gamma^{1/2}  \approx 1.29$,  throughout  the
computational domain. The  simulation domain is from $-80$  to $80$ kpc,
with $\Delta x =\Delta y=\Delta z \sim 0.13$  for the unperturbed case
and $\Delta  x =\Delta  y=\Delta z \sim  0.1$ for the  perturbed case,
which correspond  to $\sim 0.67\ {\rm  kpc}$ and $ \sim  0.5\ {\rm kpc}$
respectively.  An  important time  scale  in  the  system is  the  sound
crossing time  $\tau_{\rm s} \approx  0.78$, corresponding to  a typical
time scale $ R_{0} / C_{\rm  s0} \approx 10.0$ Myrs. Therefore, $t=1$ is
equivalent to the unit time scale $12.8$ Myrs.

In the King model we use  here, we adopt the cluster core radius $R_{\rm
c}$ to be $4.0$ (i.e., 20 kpc) and the slope $\kappa$ to be $1.0$ in the
unperturbed case  and $0.75$ in  the perturbed case.  The  injections of
magnetic flux,  mass and its associated  energies are the  same as those
described in Paper I.  The ratio between the toroidal to poloidal fluxes
of the  injected fields is characterized  by a parameter  $\alpha = 15$,
which corresponds  to $\sim  6$ times more  toroidal flux  than poloidal
flux.  The magnetic field injection  rate is described by $\gamma_b$ and
is  set  to  be  $\gamma_b=3$.  The  mass  is  injected  at  a  rate  of
$\gamma_\rho =0.1$  over a central  volume with a  characteristic radius
$r_\rho = 0.5$.  Magnetic fluxes and mass are  continuously injected for
$t_{\rm inj}  = 3.1$,  after which the  injection is turned  off.  These
parameters  correspond to  a magnetic  energy injection  rate  of $\sim
10^{43}$ ergs/s, a  mass injection rate of $\sim  0.046 M_\odot$/yr, and
an  injection  time  $\sim  40$  Myrs.   We  use  the  outflow  boundary
conditions at all outer boundaries. Note that for most of the simulation
duration, the waves and magnetic  fields stay within the simulation box,
and all magnetic fields are self-sustained by their internal currents.

\subsection{Definition of Forward and Return Currents}
As examined in  Paper II, the magnetic tower  jets can propagate without
any visible  distortion throughout  the time evolution  (see Fig.   1 of
Paper  II).  The  magnetic  tower has  a  well-ordered helical  magnetic
configuration;  a tightly  wound central  helix is  going up  inside the
tower jet and a loosely wound helix  is coming back at the outer edge of
magnetic tower  jet (see Fig.  3  of Paper II).  The  axial current flow
$J_{z}$ displays a  closed circulating current system in  which one path
flows  along  the  central  axis (the  ``forward'' current)  and  another
conically shaped  path that flows outside (the  ``return'' current) (see
Fig.   4 of  Paper II).   Here, we  define the  forward  current density
${\bolJ}^{\rm F}$  and the return  current density ${\bolJ}^{\rm  R}$ as
follows;
\begin{eqnarray} 
&&\bolJ  \equiv
\cases{{\bolJ}^{\rm  F} & ($J_{z} > 0$) \cr {\bolJ}^{\rm  R} & ($J_{z} < 0$) \cr}.
\end{eqnarray} 
Note that ${\bolJ}^{\rm F}$ (and  ${\bolJ}^{\rm R}$) is a vector and can
have  additional  radial  and  azimuthal components  $J^{\rm  F}_r$  and
$J^{\rm  F}_\phi$ ($J^{\rm R}_r$  and $J^{\rm  R}_\phi$) with  any sign.

\subsection{The Kruskal-Shafranov Criterion}  
According   to   the   well-known  Kruskal-Shafranov   (K-S)   criterion
\citep[]{S57,  K58}, a  magnetized flux  tube  (along the  $z$ axis)  is
stable to  the kink  ($m=1$) mode  as long as  the magnetic  twist angle
$\Phi(r)$ is below some critical value $\Phi_{\rm crit}$,
\begin{eqnarray}
\label{eq:K-S_limit}  
\Phi(r)  \equiv  \frac{L  B_{\phi}}{r  B_{z}}  < \Phi_{\rm   crit},   
\end{eqnarray}  
where, $L$ is  the length of the current-carrying  magnetic flux system,
$r$  is  the  cylindrical radius  of  the  flux  tube, and  $B_{z}$  and
$B_{\phi}$ are the axial and azimuthal field components in a cylindrical
coordinate,  respectively.  In  the original  K-S  criterion, $\Phi_{\rm
crit}$ is equal to $2  \pi$.  The effect of ``line-tying'', i.e., fixing
the radial motion of the magnetic fluxes at the boundary, however, could
raise the  stability threshold.  (In  the solar coronal loops,  the foot
points of  magnetic fluxes can be  anchored in a high  density region as
chromosphere.)  The  modified critical values for  a force-free magnetic
loop configuration  in solar coronal loops  are between $2  \pi$ and $10
\pi$   \citep[e.g.,][]{HP79,  EV83}.   For   the  configuration   of  an
astrophysical jet, which has  $L \gg r$, inequality (\ref{eq:K-S_limit})
can be  easily violated.   If the MHD  jets originate from  an accretion
disk, the bottom part  of jets may be line-tied in the  disk and the top
part of  jets can be  free against the  radial motion; i.e.,  a ``semi''
line-tying configuration.   (Further argument of  the stabilizing effect
by semi line-tying  will be presented in \S 4.2.)   For lack of detailed
linear analysis of ``semi'' line-tying magnetic configurations, we still
define $\Phi_{\rm  crit} = 2  \pi$, which should  only be regarded  as a
guide for gauging the stability, rather than a precise threshold.  Here,
we replace $L$ by a  spectrum of axial wavelengths $\lambda$, and define
the critical wavelength to be the one that is marginally stable:
\begin{eqnarray}   
\label{eq:K-S_crit_wavelength}   
\lambda_{\rm crit}   \equiv \frac{\Phi_{\rm crit}   
r   B_{z}}{B_{\phi}}   =   \frac{2   \pi   r B_{z}}{B_{\phi}}.      
\end{eqnarray} 
The K-S stability criterion (eq. [\ref{eq:K-S_limit}]) then becomes
\begin{eqnarray}  
\lambda < \lambda_{\rm crit}, 
\end{eqnarray} 
so that wavelengths greater  than $\lambda_{\rm crit}$ could be unstable
to the CDI kink mode.

\subsection{Diagnostics of Non-axisymmetric Mode}
To better diagnose the non-axisymmetric distortion, we perform the power
spectrum analysis of the magnitude of the current density $|\bolJ|$.  We
analyze   the   modal  structure   of   $|\bolJ|$   by  performing   the
volume-averaged Fourier transform:
\begin{eqnarray}
\label{eq:fourier_1} 
f(m,\,k) = \frac{1}{V_{\rm cl}}
\int\int\int_{V_{\rm cl}} |\bolJ| e^{i(m\phi+kz)}\,r\,dr\,d\phi\,dz,
\end{eqnarray} 
where  $|\bolJ|  \equiv  \bigl(J_r^{2}+J_{\phi}^{2}+J_z^{2}\bigr)^{1/2}$
and $V_{\rm cl}$ is the  cylindrical volume that encloses $\bolJ (\equiv
\bolJ^{\rm F}+\bolJ^{\rm R})$
\begin{eqnarray} 
\label{eq:fourier_2} 
V_{\rm cl}= \int_{r_{\rm a}}^{r_{\rm b}} \int_{0}^{2 \pi} 
\int_{z_{\rm a}}^{z_{\rm b}}\,r\,dr\,d\phi\,dz.  
\end{eqnarray} 
We use  the values $r_{\rm  a}=0.0$, $r_{\rm b}=7.75$,  $z_{\rm a}=0.0$,
and $z_{\rm b}=15.5$  in the ``upper'' volume and  use $z_{\rm a}=-15.5$
and $z_{\rm b}=0.0$ in the  ``lower'' volume. The quantity $f(m,\,k)$ is
a function of  the azimuthal mode number $m$ and  axial wave number $k=2
\pi/ \lambda$ (corresponding to a characteristic wave length $\lambda$),
and it is a function of time as well.

Finally, we identify the power spectrum as 
\begin{eqnarray}
\label{eq:fourier_3}
&&|f(m,\,k)|^{2}=
\left\{{\rm Re} \bigl[f(m,\,k) \bigr] \right\}^{2} +
\left\{{\rm Im} \bigl[f(m,\,k) \bigr] \right\}^{2},
\end{eqnarray}
\begin{eqnarray}
\label{eq:fourier_4}
&&{\rm Re} \bigl[f(m,\,k) \bigr] \nonumber \\
&&\quad =\frac{1}{V_{\rm cl}} \int\int\int_{V_{\rm cl}} 
|\bolJ| \cos\,(m\phi+kz)\,r\,dr\,d\phi\,dz,\\
\label{eq:fourier_5}
&&{\rm Im} \bigl[f(m,\,k) \bigr] \nonumber \\
&&\quad =\frac{1}{V_{\rm cl}} \int\int\int_{V_{\rm cl}} 
|\bolJ| \sin\,(m\phi+kz)\,r\,dr\,d\phi\,dz.
\end{eqnarray}
We will examine the  time-dependent behavior of the  power spectrum
and the growth of each mode.

\section{RESULTS}
\subsection{Unperturbed Case --- Dynamical Stabilization against the CDI Kink Mode}
\label{sec:A} 
\begin{figure}
\begin{center}
\includegraphics[scale=0.95, angle=0]{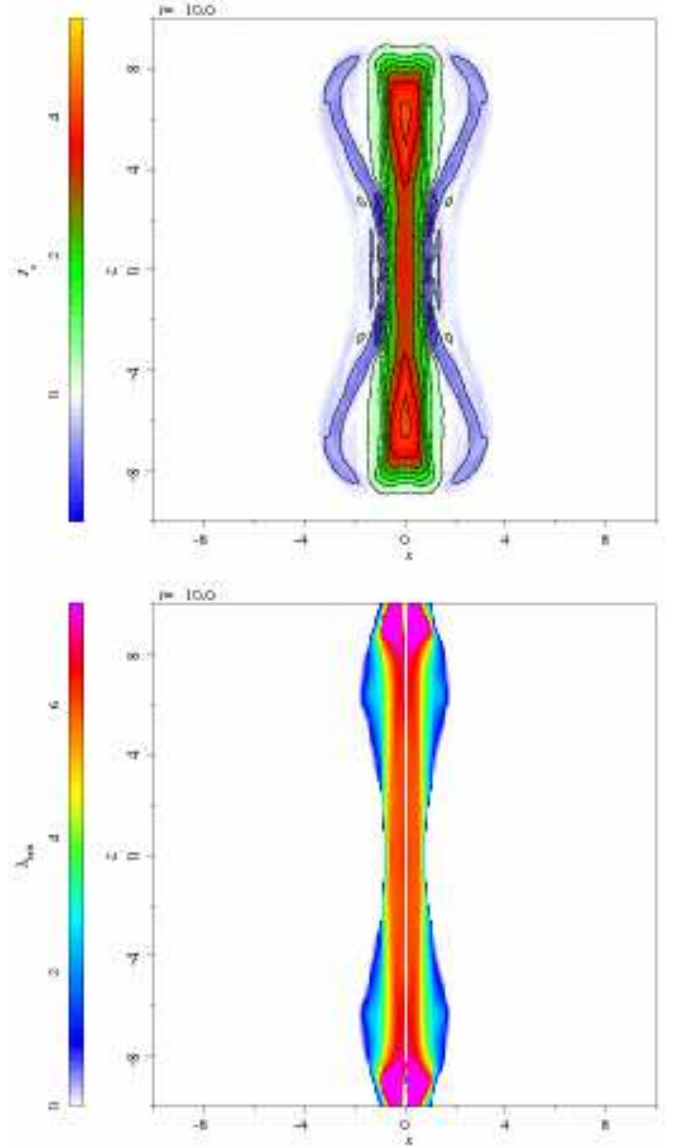}
\caption{\label{fig:t7.5_Jz_WLc}  {\em Top}:  Distribution of  the axial
current  density  $J_{z}$  in  the  $x-z$  plane at  $t=10.0$    in  the
unperturbed case (Paper II).  {\em Bottom}: Distribution of the critical
wavelengths  $\lambda_{\rm  crit}$  at  $t=10.0$  in  the  $x-z$  plane,
corresponding to the original  Kruskal-Shafranov criterion for CDI $m=1$
mode in the unperturbed case (Paper II).}
\end{center}
\end{figure}

We first re-examine the numerical result of Paper II as an
unperturbed case,  from a viewpoint of stability.   For that simulation,
we did not impose any  initial perturbations to the background state. So
the only  available perturbations  are from the  numerical noise  in the
double precision  computations.  Figure \ref{fig:t7.5_Jz_WLc}  shows the
snapshots of  $J_z$ ({\em top}) and $\lambda_{\rm  crit}$ ({\em bottom})
in the final  stage ($t=10.0$).  Again, we can  identify the circulating
path of  axial current flow $J_{z}$  which consists of  both the forward
axial  current density  $J_{z}^{\rm F}~(J_z>0)$  and the  return current
$J_{z}^{\rm  R}~(J_z<0)$ in the  {\em top}  panel ($J_{z}^{\rm  F}$ lies
around  the central  axis, while  $J_{z}^{\rm R}$  surrounds $J_{z}^{\rm
F}$).  The  maximum strength  of $|J_{z}^{\rm F}|$  is much  larger than
that of $|J_{z}^{\rm R}|$, indicating that a tightly wound central helix
is going  up at the  center of  the tower and  a loosely wound  helix is
coming back at the tower edge.  The paths of forward and return currents
are approaching one another around the mid-plane $|z| \lesssim 4.0$.  On
the other hand, the path of  return current is radially expanding at the
upper axial position $|z| \gtrsim  4.0$.  This causes a {\em separation}
of current  flow paths in  a thermally confined  MHD jet with  a helical
magnetic field. (This  plays an important role in  the excitation of the
external kink mode and it will be discussed in \S 4.3.)

\begin{figure*}
\centerline{{\includegraphics[scale=0.9, angle=0]{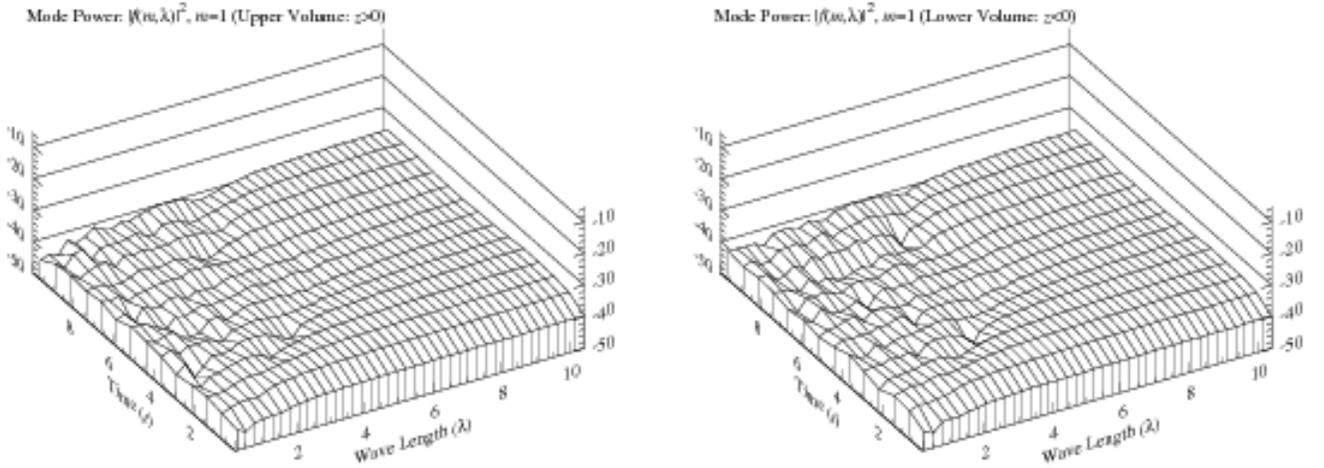}}}
\caption{\label{fig:mode_power_p19} Space-time ($\lambda$,\,$t$) diagram
of the azimuthal Fourier power (a  natural log scale) for CDI mode $m=1$
in the unperturbed case (Paper  II). {\em Left}: For the upper volume $z>0$.
{\em Right}: For the lower volume $z<0$.  }
\end{figure*}

\begin{figure}
\begin{center}
\includegraphics[scale=0.95, angle=0]{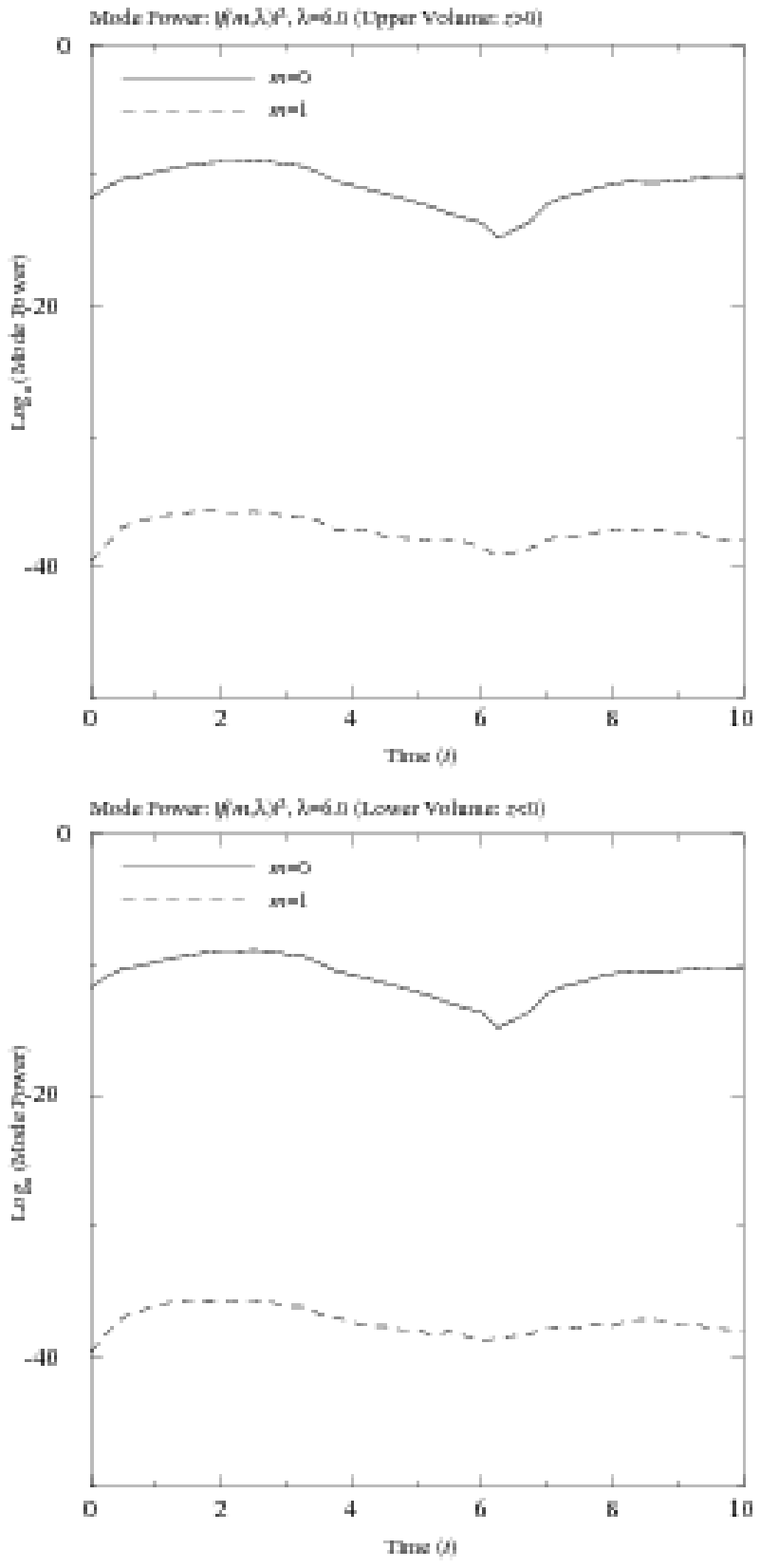}
\caption{\label{fig:growth_rate_p19} The time variation of the azimuthal
Fourier power  (a natural log scale)  in modes $m=0,\,1$  with a specified
wave length $\lambda=6.0$ in the unperturbed case (Paper II). {\em Top}:
For the upper volume $z>0$. {\em Bottom}: For the lower volume $z<0$.  }
\end{center}
\end{figure}

As seen in the {\em bottom}  panel, $\lambda_{\rm crit}$ has a range 2.0
$\sim$  6.0 in  the  core part  of  $J_{z}^{\rm F}$,  and  has no  axial
dependency except at the tower front.  While $\lambda_{\rm crit}$ has no
range in any part of $J_{z}^{\rm  R}$, meaning that the CDI kink mode is
unlikely to grow at the return current region.  Taking the whole part in
the  axial   direction  into   consideration,  the  maximum   length  of
current-carrying column is $L \sim 16$ at $t=10.0$, which is larger than
the  distributed $\lambda_{\rm crit}$  in the  core part  of $J_{z}^{\rm
F}$.  However, the non-axisymmetric  displacement does not seem to occur
as seen in  both panels.  We next show the result  of the power spectrum
analysis of $\bolJ^{\rm F}$ in order to confirm the non-existence of the
growing kink mode.  Figure \ref{fig:mode_power_p19} shows the space-time
$(\lambda, t)$ diagrams of  the power spectrum $|f(m, \lambda)|^{2}$ for
$m=1$  in  the  unperturbed  case.    We  can  not  detect  any  growing
non-axisymmetric mode ($m=1$) over the  wider range of axial wave length
($\lambda=0.26 \sim  10.42$).  The time variation of  the power spectrum
of  $f(m, \lambda)$  with a  selected wavelength  $\lambda=6.0$  in Fig.
\ref{fig:growth_rate_p19}.  Throughout the  time evolution, the power in
$m=1$ mode is  always below the axisymmetric ($m=0$)  one. Note that the
variation of the $m=0$ mode  (seen in both volumes around $t=4\sim8$) is
due  to  the axisymmetric  deformation  of  the  magnetic tower  into  a
slender-shaped structure by the gravitational contraction radially.

\begin{figure}
\begin{center}
\includegraphics[scale=1.1, angle=0]{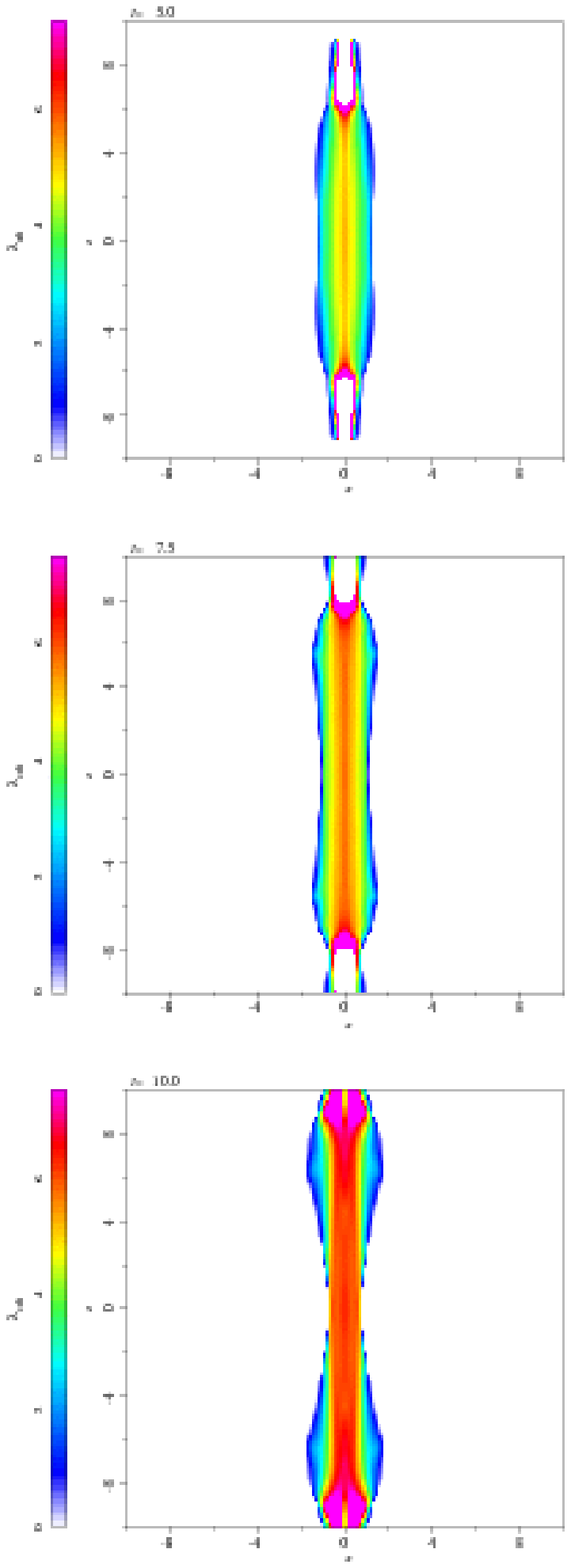}
\caption{\label{fig:WLcrit_evo}  Time evolution  of the  K-S critical
wavelength  $\lambda_{\rm crit}$  in the  $x-z$ plane  at  $t=5.0$ ({\em
Top}), 7.5  ({\em Middle}), and  10.0 ({\em Bottom}) in  the unperturbed
case (Paper II).  }
\end{center}
\end{figure}

It may seem puzzling that,  for the unperturbed case, the magnetic tower
jet is stable  even with $L > \lambda_{\rm crit}$.   Here, we consider a
possible  mechanism of the  stabilizing effect  in our  results.  Figure
\ref{fig:WLcrit_evo}  shows  the  time  evolution of  the  K-S  critical
wavelength  $\lambda_{\rm crit}$  in  the unperturbed  case.  A  gradual
increase   of   the   distributed   $\lambda_{\rm  crit}   (=2   \pi   r
B_{z}/B_{\phi})$   in   the   forward  axial   current-carrying   column
$J_{z}^{\rm  F}$  can  be  observed.   This corresponds  directly  to  a
decrease  of  the  ratio  $B_{\phi}/B_{z}$.  Because  the  injection  of
magnetic fluxes  into the central volume  in the system  has been turned
off  at $t =  3.1$, the  twisted magnetic  fluxes are  naturally relaxed
during the expansion of the magnetic tower.  The threshold length of the
marginally  stable  of $J_{z}^{\rm  F}$  in  the  axial ($z$)  direction
becomes  longer   as  the  magnetic  tower   proceeds.   Therefore  this
``dynamical relaxation''  of magnetic twists plays an  important role in
the stabilization in our current results.  We will discuss other effects
on the stabilization in \S 4.2.

\subsection{Perturbed Case --- Growth of the CDI External/Internal Kink Modes}
\label{sec:B}  We next  examine  the non-axisymmetric  evolution of  the
magnetic  tower  jet in  a  gravitationally  stratified atmosphere.   To
excite  non-axisymmetric modes  ($m  \ge 1$),  a  small random  velocity
perturbation (4\% of initial sound speed) has been added into the system
as an initial condition.

\begin{figure*}
\centerline{{\includegraphics[scale=1.2, angle=-90]{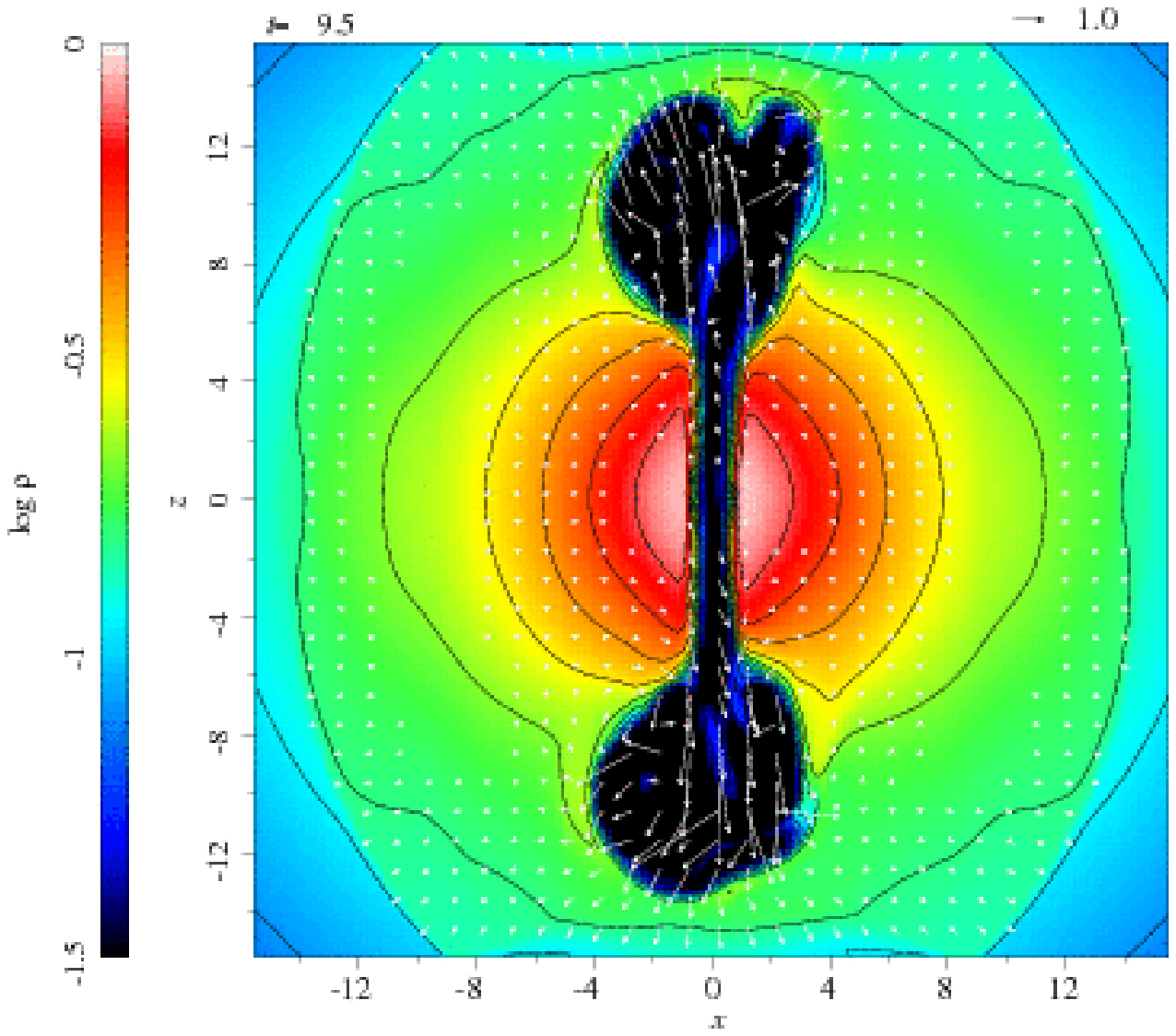}}}
\caption{\label{fig:de_final}   Distribution  of   the   density  $\rho$
(logarithmic scale) in the $x-z$  plane is shown along with the poloidal
velocity field (arrows)  at $t=9.5$ in the perturbed  case. Similar to
the unperturbed case (Paper II),  density cavities are formed due to the
magnetic field expansion, while the non-axisymmetric motion can be seen.
The length of the arrow at the upper right shows the unit speed.  }
\end{figure*}

Figure \ref{fig:de_final} shows a snapshot  of the density $\rho$ at the
final stage  $t=9.5$ (see  Fig.  1  of Paper II  for comparison)  in the
perturbed case.  Even though the  configuration of the magnetic tower is
non-axisymmetric, the  main features  remain similar to  the unperturbed
case.  In general, the structure  of the magnetic tower consists of both
a well-collimated ``body'' and radially extended ``lobes''.  Several key
features  in the  magnetic  tower  jets (the  MHD  wave structures,  the
heating  process  at  the  tower  front, the  cylindrical  radial  force
equilibrium at the tower  edge, and dynamically collimating process) are
the  same as  the  unperturbed case  (see  Paper II  for details).   The
Alfv\'en speed becomes  large (about three times of  local sound speed),
while the plasma $\beta  (\equiv 2p/B^2)$ becomes small ($\beta \lesssim
0.1$)  inside the  density cavities  due  to the  expansion of  magnetic
fluxes.   The bulk  flow  inside  the magnetic  tower  is sub-sonic  and
sub-Alfv\'enic, while the fronts of the preceding hydrodynamic shock and
the magnetic tower can propagate super-sonically.

\begin{figure*}
\centerline{{\includegraphics[scale=1.1, angle=0]{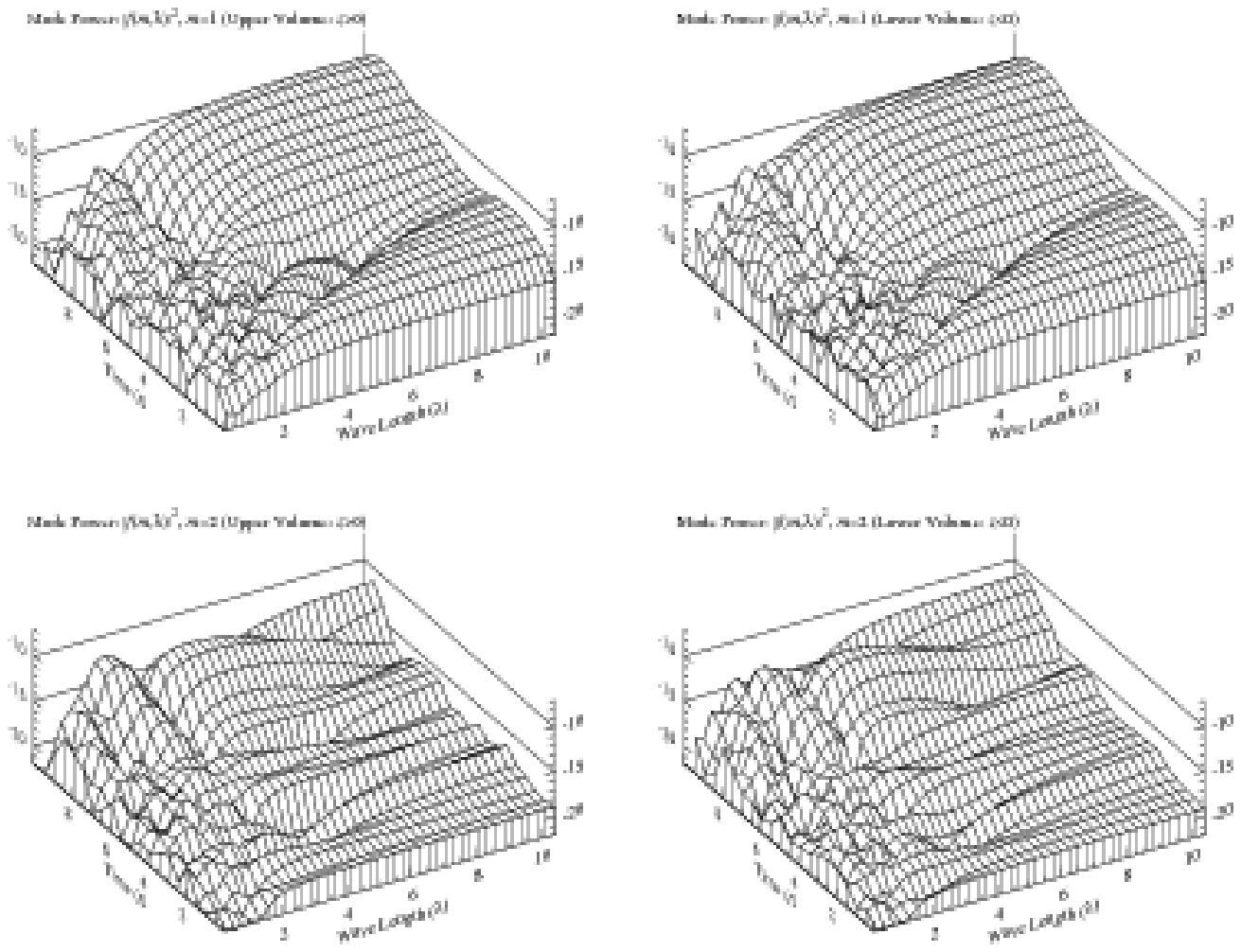}}}
\caption{\label{fig:mode_power_b3}               Similar              to
 Fig. \ref{fig:mode_power_p19},  but for modes  $m=1$ ({\em Top})  and 2
 ({\em Bottom})  in the  perturbed case.  {\em  Left}: For the upper volume
 $z>0$. {\em Right}: For the lower volume $z<0$.  }
\end{figure*}

\begin{figure}
\begin{center}
\includegraphics[scale=0.95, angle=0]{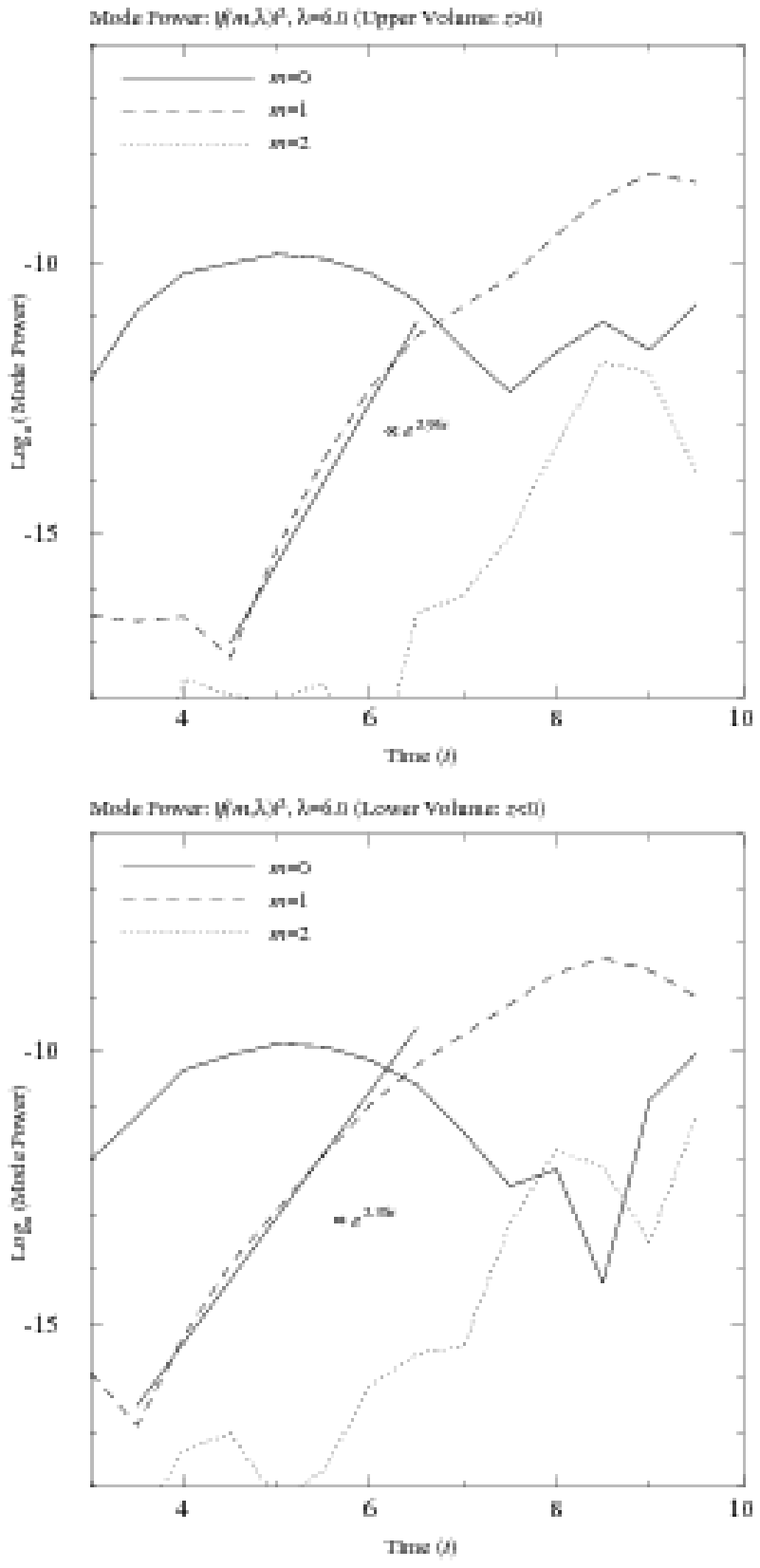}
\caption{\label{fig:growth_rate_b3}              Similar              to
 Fig. \ref{fig:growth_rate_p19}, but for  modes $m=0-2$ in the perturbed
 case. Straight  lines are  derived by fitting  the slopes of  the $m=1$
 mode  growth  to  a  linear  function.  {\em  Top}:  For the upper  volume
 $z>0$. {\em Bottom}: For the lower volume $z<0$.  }
\end{center}
\end{figure}

Similar  to  Fig. \ref{fig:mode_power_p19},  we  present the  space-time
$(\lambda, t)$ diagrams of the power spectrum $|f(m, \lambda)|^2$ in the
perturbed     case     for    the     $m=1$     and     2    in     Fig.
\ref{fig:mode_power_b3}. Clearly,  the growth of  the $m=1$ mode  can be
seen over a wide  range of $\lambda$ after $t \sim 4$  in both upper and
lower volumes.  Furthermore, the $m=2$ (elliptical) mode also grows at a
relatively short  wavelength around $\lambda =  2 \sim 3$  after $t \sim
6$.  To identify  the growth rates of different modes,  we show the time
variation of the power spectrum of $f(m, \lambda)$ in the perturbed case
with     a     selected     wavelength     $\lambda=6.0$     in     Fig.
\ref{fig:growth_rate_b3}  (similar to  Fig.  \ref{fig:growth_rate_p19}).
The  $m=1$ mode  in  both volumes  exhibits  an approximate  exponential
growth on a dynamical timescale during  $t= 4 \sim 6$.  At later stages,
after the $m=1$ mode grows above the axisymmetric $m=0$ mode, its growth
rate  decreases  slightly,  but  still keeps  growing  exponentially  to
saturation at  the final stage.  The  $m=2$ mode also  shows the similar
behavior with the $m=1$ mode at  the earlier stage around $t= 6 \sim 8$,
but it saturates at or below the level of the $m=0$ mode.

A  linear fit  using a  least-square method  has been  performed  on the
initially  exponential growth  for the  $m=1$ mode  to extract  a linear
growth rate
\begin{eqnarray}
{\rm Im}\,(\omega) \sim \frac{d\,{\rm ln}\,|f(m, \lambda)|^{2}}{dt}.
\end{eqnarray}
The  estimated growth  rates in  the perturbed  case are  also  given in
Fig.  \ref{fig:growth_rate_b3}.  The  growth  rate in  the upper  volume
appears to be a bit stronger  than that in the lower volume.  We confirm
the  distributed   Alfv\'en  speed  $V_{\rm  A}$  in   the  surfaces  of
$\bolJ^{\rm F}$ is of the order of $8 - 9$ in our normalized unit during
the   time  evolution.    The  spatial   length  scale   $\xi$   of  the
non-axisymmetric ($m > 0$) distortions in the radial direction is of the
order of  $2 - 4$, as  seen in Fig.  \ref{fig:de_final}.  Therefore, the
inverse  of  the  Alfv\'en  crossing  time $\tau_{\rm  A}^{-1}$  in  the
distorted part of magnetic tower jet is given by
\begin{eqnarray}
\tau_{\rm A}^{-1}=\frac{V_{\rm A}}{\xi} \sim 2.0 - 4.5.
\end{eqnarray}
This is, in general, consistent with the timescales of the growing $m=1$
mode derived  from Fig. \ref{fig:growth_rate_b3}.  So,  our results lead
us  to conclude  that the  non-axisymmetric distortion  in  the magnetic
tower jet is caused dominantly by the normal current-driven kink ($m=1$)
mode.

\begin{figure}
\begin{center}
\includegraphics[scale=0.95, angle=0]{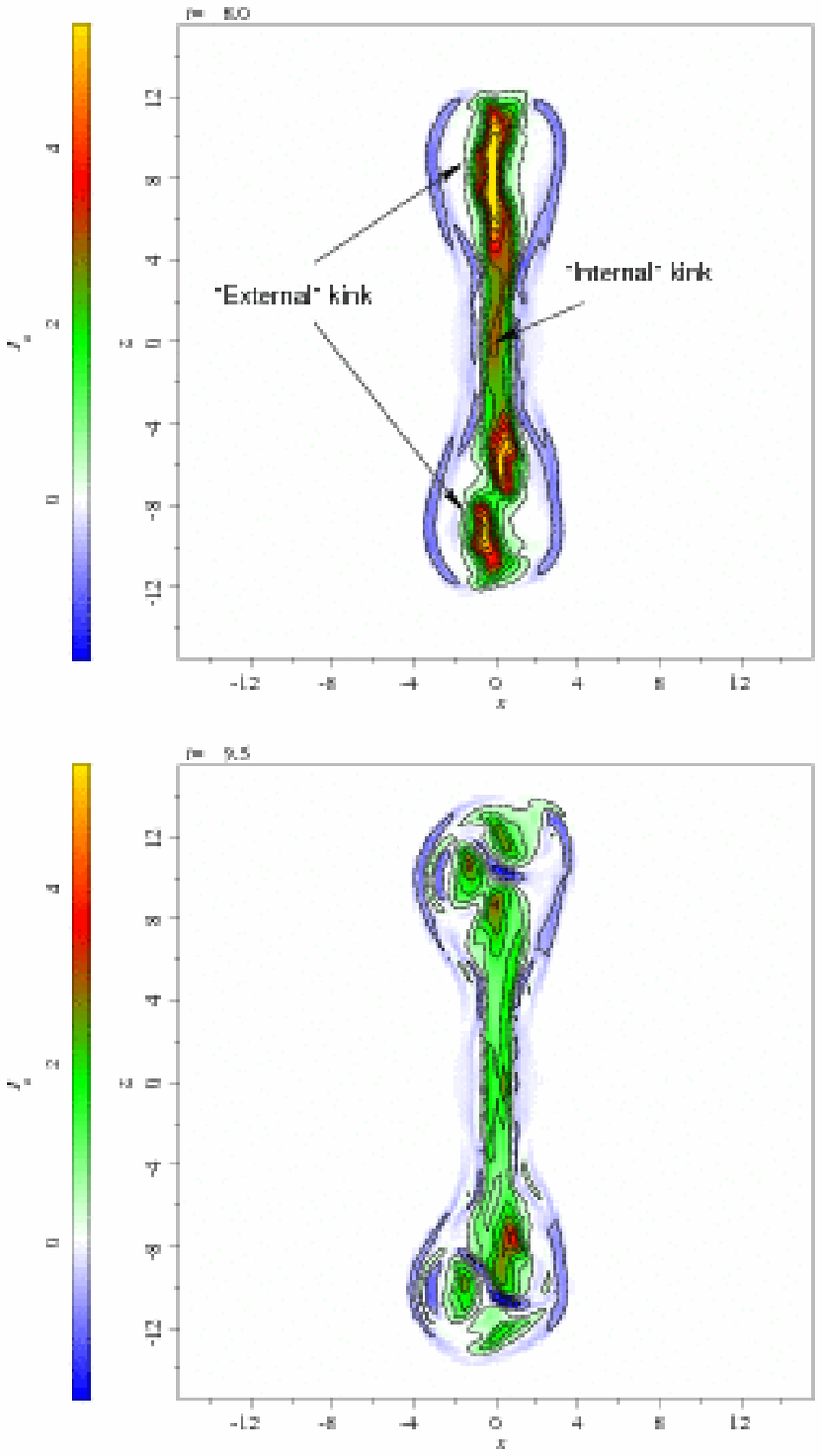}
\caption{\label{fig:Jz_evo}  Snapshots  of  the  axial  current  density
$J_{z}$ in the $x-z$ plane at $t=8.0$ ({\em Top}) and 9.5 ({\em Bottom})
in   the   perturbed   case.    Two   types  of   growing   kink   modes
(``External''/''Internal'') are illustrated on the {\em Top} panel.  }
\end{center}
\end{figure}

\begin{figure}
\begin{center}
\includegraphics[scale=0.3, angle=0]{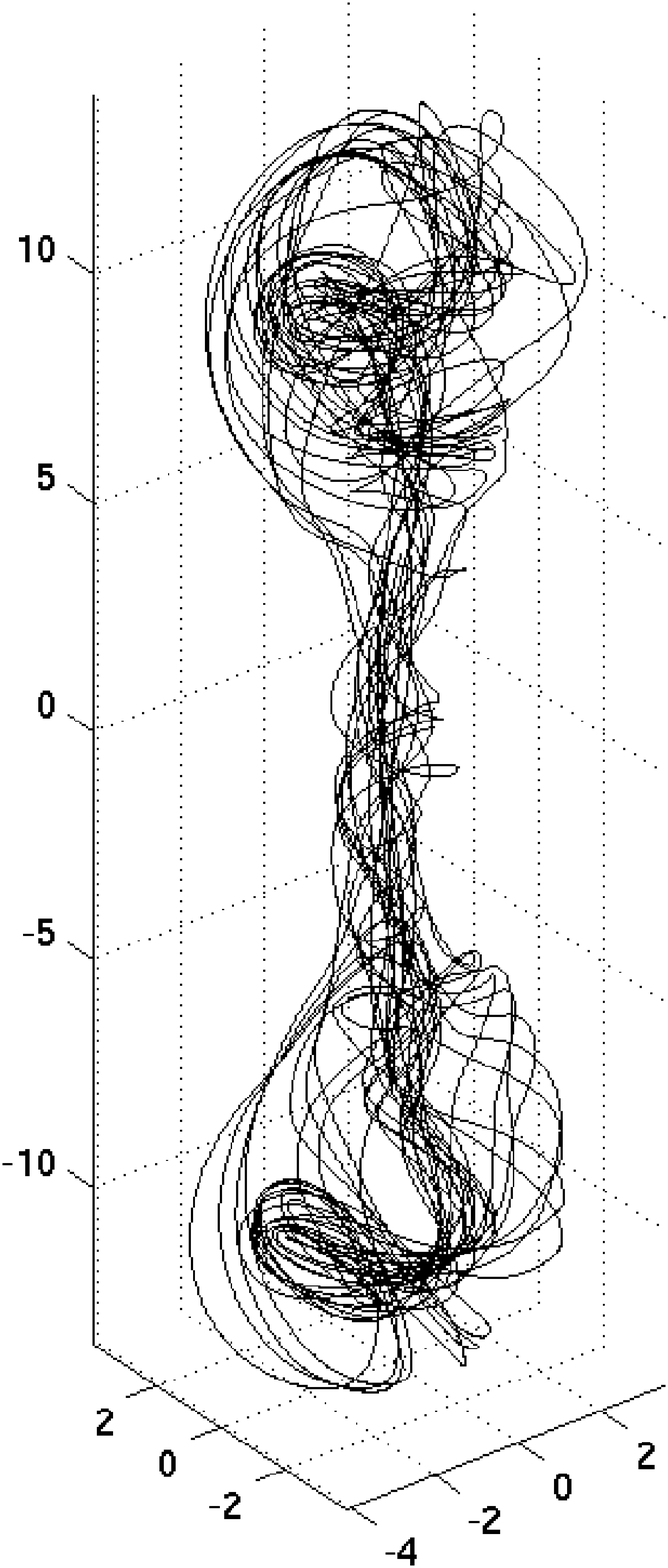}
\caption{\label{fig:3Dlines}  Three-dimensional  view  of some  selected
magnetic lines at $t=9.5$ in the perturbed case.  }
\end{center}
\end{figure}

Two types  of CDI  modes are identified  in the context  of magnetically
controlled fusion  plasma; the  ``external'' and the  ``internal'' modes
\citep[]{BA78}.   The external  mode  can be  identified  with a  radial
distortion, while in the internal  mode, there is no radial perturbation
at the outer  edge of the current-carrying column.   One of our findings
in  the  perturbed magnetic  tower  jet is  that  the  CDI external  and
internal  modes  ``co-exist'' during  the  dynamical evolution.   Figure
\ref{fig:Jz_evo}  shows snapshots of  $J_{z}$ in  the perturbed  case at
$t=8.0$ and 9.5.  As seen in  the {\em top} panel ($t=8.0$), the edge of
$\bolJ^{\rm  F}$ is  clearly shifted  to  an offset  axial direction  at
higher latitude ($|z|  \gtrsim 4$).  On the other  hand, that edge never
shifts  non-axisymmetrically  at  lower  latitude  ($|z|  \lesssim  4$).
However,  the non-axisymmetric  distribution of  $\bolJ^{\rm F}$  at the
inner  part  can be  detected,  indicating  that  the internal  mode  is
certainly growing.  At  the final stage ($t=9.5$: {\em  bottom}), we can
see the well-grown  wiggles inside the ``lobes'', while  the edge of the
``body''  part in  the magnetic  tower still  remains quasi-axisymmetric
except   for   the   inner    part.    Taking   the   result   of   Fig.
\ref{fig:mode_power_b3}  into  consideration,  the  $m=1$ kink  mode  is
dominant in  the external CDI at  the lobes, while  the $m=2$ elliptical
mode may  be dominant in  the internal CDI  at the body of  the magnetic
tower.  In addition, we illustrate  a 3-D view of some selected magnetic
lines  of force  in the  perturbed case  at the  final stage  $t=9.5$ in
Fig. \ref{fig:3Dlines}. It indicates that a tightly wound central helix,
which goes up along the central axis, relaxes in the lobe part where the
external  kink mode  has taken  place, and  a large-scale  loosely wound
twist is formed there.

\section{DISCUSSION}
\label{sec:DS}
\subsection{Suppression of Shear-driven Instability}
We first  discuss the possibility of  the interplay between  CDI and KHI
modes  in  our results.  The  examinations  of  KHI in  linear/nonlinear
regimes  have been  extensively  performed  by P.   E.   Hardee and  his
collaborators \citep[see, {\it e.g.},][and references therein]{H04}.  In
general, non-axisymmetric ``surface'' ({\em external}) modes ($m>0$) are
stable   against    KHI   for   sub-Alfv\'enic    flow.    However,   in
super-Alfv\'enic but trans-fast magnetosonic flow, they can be unstable.
Only   axisymmetric   surface   mode   ($m=0$)   remains   unstable   in
sub-Alfv\'enic  flow,  although  with  a relatively  small  growth  rate
\citep[]{B89}.  A potential zone  of unstable region in non-axisymmetric
KHI surface modes should exist in downstream of the Alfv\'en surface.

The instability criterion for  non-axisymmetric KHI surface modes is
as follows \citep[]{HR02} :
\begin{eqnarray}
\label{eq:KHIs}
\Delta  V > V_{\rm  A s}  = \sqrt{\frac{\rho_{j}+\rho_{e}}{4 \pi
\rho_{j}\,\rho_{e}} (B_{j}^{2}+B_{e}^{2})}
\end{eqnarray}
where $\Delta V \equiv |V_{j}-V_{e}|$  is the velocity shear and $V_{\rm
A s}$ is the {\em  surface} Alfv\'en speed (subscript $j$ corresponds to
the  jet itself and  $e$ corresponds  to the  external medium).   On the
other hand,  the non-axisymmetric ``body'' (internal)  modes ($m>1$) can
become  important   and  affect  the  jet  interior   in  the  following
situations: (1) if the  jet velocity exceeds the super-fast magnetosonic
speed $V_{f}$:
\begin{eqnarray}
\label{eq:body1}
V_{j} > V_{f},
\end{eqnarray}
or (2) if the flow velocity is slightly below the slow magnetosonic
speed $V_{s}$: 
\begin{eqnarray}
\label{eq:body2}
\frac{C_{s}V_{\rm A}}{\sqrt{C_{s}^2+V_{\rm A}^2}}< V_{j} < V_{s},
\end{eqnarray}
where $C_{s}$ is the sound speed and $V_{\rm A}$ is the Alfv\'en speed 
\citep[]{HR99}.

\begin{figure*}
\centerline{{\includegraphics[scale=1.2, angle=0]{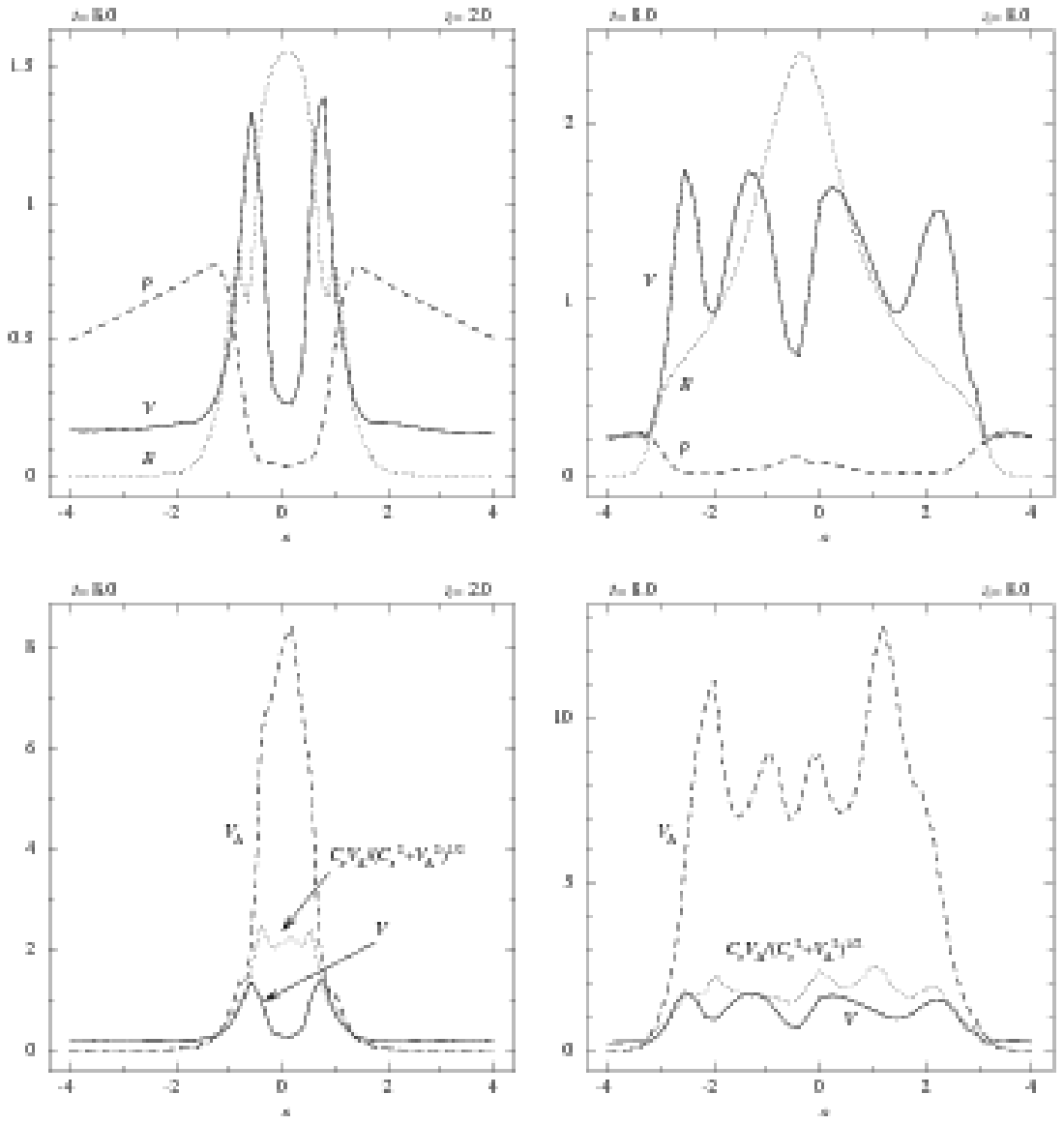}}}
\caption{\label{fig:KHIstable} Transverse profiles  in the $x$ direction
of several quantities  at $t=8.0$ on $z=2.0$ ({\em  Left} panels) and on
$z=8.0$ ({\em Right} panels) in  the perturbed case.  The bulk speed $V$
($=\sqrt{V_x^2+V_y^2+V_z^2}$),  the  density  $\rho$, and  the  magnetic
field   strength  $B$   ($=\sqrt{B_x^2+B_y^2+B_z^2}$)   are  shown   for
inspecting the KHI surface modes ({\em Top} panels). The bulk speed $V$,
the  Alfv\'en   speed  $V_{\rm  A}$,   and  the  quantity   $C_s  V_{\rm
A}/(C_s^2+V_{\rm  A}^2)^{1/2}$ are  shown  for inspecting  the KHI  body
modes ({\em Bottom} panels).}
\end{figure*}

Figure  \ref{fig:KHIstable} shows  the  transverse profiles  in the  $x$
direction of several  quantities at $t=8.0$ in the  perturbed case.  Two
different axial positions $z=2.0$ and 8.0 are selected for examining the
KHI surface/body modes. For the surface modes, we inspect the transverse
distribution of  the bulk  flow speed $V$,  the density $\rho$,  and the
magnetic  field strength  $B$.  Two  distinct velocity  shears  (in both
positive and negative  $x$ regions) are identified in  the $x$ direction
on $z=2.0$ ({\em upper-left}); an  outer part $0.6 \lesssim |x| \lesssim
1.5$ and  an inner part $0.2  \lesssim |x| \lesssim 0.6$.   At the outer
part, $\Delta  V \sim 1.2$  and $V_{\rm As}  \sim 1.8$ and at  the inner
part, $\Delta  V \sim  1.1$ and $V_{\rm  As} \sim 8.7$.   Therefore, the
inequality $\Delta V < V_{\rm  As}$ is satisfied.  Similarly, we can see
many distinct velocity shears  in the {\em upper-right} panel.  However,
even in these shears, we confirm  that the inequality $\Delta V < V_{\rm
As}$ still holds everywhere.  This  indicates that the KHI surface modes
are  completely stabilized,  because  there is  no significant  velocity
shear exceeding the local surface Alfv\'en speed in the transverse ($x$)
direction.  For  the body modes, we inspect  the transverse distribution
of the  bulk flow speed $V$,  the Alfv\'en speed, and  the quantity $C_s
V_{\rm A} / (C_s^2+V_{\rm  A}^2)^{1/2}$.  Inside the magnetic tower, the
plasma $\beta$  ($= 2p/B^2$) is always  less than unity  during the time
evolution. Therefore, the phase speed of the MHD fast mode wave is equal
(in  the local  magnetic  field  direction) or  larger  (in the  oblique
direction of the  local magnetic field) than the  Alfv\'en speed, $V_{f}
\ge  V_{\rm  A}$.   As  seen   in  both  {\em  Bottom}  panels  in  Fig.
\ref{fig:KHIstable},   the   bulk  flow   speed   is  always   extremely
sub-Alfv\'enic, and  furthermore, the  inequality $V <  C_s V_{\rm  A} /
(C_s^2+V_{\rm A}^2)^{1/2}  $ is also  satisfied in the ``body''  part of
the  magnetic tower  jet. These  indicate that  the KHI  body  modes are
suppressed as well.  We conclude that the KHI modes play no role and the
normal CDI modes create the non-axisymmetric structures in our results.

\subsection{Stabilization/Destabilization of Magnetic Tower Jets} 
As examined in \S \ref{sec:A}, the unperturbed magnetic tower jet can be
stabilized against the  CDI kink mode beyond the  K-S criterion.  One of
the certain mechanisms  of the stabilizing effect, which  we can identfy
in our unperturbed case, is the dynamical relaxation of magnetic twists.
However, this  process may  contribute the {\em  marginal stabilization}
under an ideal situation (no perturbation except for the numerical noise
is  assumed  as an  initial  condition).   If  the system  is  perturbed
randomly  with a  finite amplitude  (a few  percent of  the  local sound
speed),  then the  non-axisymmetric  ($m \ge  1$) internal/external  CDI
modes  can  grow.   The   disturbances  in  the  system  are  reasonable
(especially   when  related   to  possible   astrophysical  situations).
Therefore, the magnetic tower jet  in our treatments, will eventually be
subject to CDI.  Here, we give further discussions on the stabilities of
magnetic tower jets.

\subsubsection{Line-tying Effect} 
In  the  solar  coronal  magnetic  loops,  it is  well  known  that  the
line-tying  effect  at the  foot  points  (chromosphere)  can raise  the
original K-S stability threshold (dense materials can prevent the radial
motion  of  twisted  magnetic  fluxes).   How  about  the  case  in  the
astrophysical  jets?  Anchored  magnetic  fluxes to  the accretion  disk
and/or  stellar/black hole  surfaces may  be considered  as a  semi (one
side) line-tying  boundary condition in  a certain dynamical  stage.  An
important clue  to this  can be derived  from laboratory  experiments of
magnetic  towers  \citep[]{HB02,   HB03}.   Their  quasi  ``force-free''
magnetic  towers can exhibit  the systematic  $m=1$ distortion  when the
current-carrying system  satisfies the original K-S limit.   It might be
unlikely that such  a semi line-tying effect can  stabilize the jet even
in  the astrophysical  situation.  In  our global  modeling  of magnetic
tower jets,  we do  not solve  the evolution of  the central  region (an
accretion disk, etc.).   Therefore, we can not apply  it directly to the
AGN central region.  It is still an open issue whether a semi line-tying
effect can work or not in the real astrophysical jets.

\subsubsection{Internal Thermal Pressure}
\label{sec:DS_ITP}   As   discussed  in   Paper   II,  the   ``forward''
current-carrying  column  $\bolJ^{\rm  F}$  never reaches  a  force-free
equilibrium.   Instead, it  will be  supported by  the outward-directed,
local (core of the magnetic towers) pressure gradient force: $- \nabla p
+ \bolJ \times \bolB\simeq0$ during  a dynamical evolution (see Fig.  10
in  Paper II).   In terms  of minimization  of the  MHD  energy integral
\citep[]{B58}, a finite  gas pressure in the system  (plasma $\beta \neq
0$) can contribute to the  stabilization.  However this can be effective
when $\beta \gtrsim 1$, i.e., the gas pressure is more dominant than the
magnetic pressure \citep[for  $m=1$ kink mode, see][]{HP79}.  Therefore,
we may not expect a strong stabilization by the pressure in the magnetic
tower  ($\beta \lesssim  0.1$).  Furthermore,  the  outward-directed gas
pressure gradient  itself ($-  \nabla p >  0$) actually  destabilizes the
tower through  the PDI (interchange  and/or ballooning) modes,  which is
driven  by  the  perpendicular   currents  (in  contrast  to  the  axial
[parallel]  currents, considered  mainly in  this paper).   We, however,
could not  find any  growth of  the PDI modes,  in both  the unperturbed
(Paper II) and the perturbed cases, implying no direct contribution
of the  pressure gradient  to the destabilization  by the PDI  modes (see
next section  for more  details).  Also, we  can not confirm  any direct
contributions of  the internal thermal pressure effects  ($\beta \neq 0$
and $-  \nabla p$) to  the destabilization by the  CDI internal/external
modes.

\subsubsection{Magnetic Field Configuration}  
\label{sec:DS_RFP} We next consider  the magnetic field configuration in
magnetic  tower jets.   In  the context  of  magnetic controlled  fusion
systems,  the helical  field configuration  in  the tower  model can  be
identified as the reversed field pinch (RFP) profile as seen in Fig.  11
in Paper II.  The axial component  $B_{z}$ has a crucial shear (its sign
reverses from ``+'' to ``-'')  in the radial direction \citep[RFP in the
context of astrophysical  jets, see also][]{B06}.  Several stabilization
effects in  RFP are  summarized in \citet[]{F82};  (i) a hollow  or very
flat  pressure  profile ($\nabla  p\simeq  0$)  are  required near  the
central ($z$)  axis to suppress the PDI  sausage (interchange, internal)
mode ($m=0$), but  (ii) it can be suppressed  when $\beta \lesssim 0.5$.
This may be the situation in the present paper.  As already discussed in
the previous  section (\S \ref{sec:DS_ITP}), a  finite pressure gradient
exists to  contribute to the radial  force equilibrium in  the core part
(around the central axis) and this  might play a role in driving the PDI
modes.   However, the plasma  $\beta$ is  small ($\beta  \lesssim 0.1$),
indicating that the  destabilization by the PDI modes  may be suppressed
due to the  small plamsa $\beta$.  Also, (iii)  the strong $B_{z}$ shear
can prevent the destabilization  by the CDI internal/external kink modes
($m=1$).   In our  results,  this shear  may  provide the  stabilization
against  the  CDI external  modes,  while it  does  not  affect the  CDI
internal modes at the lower latitude  $|z| \lesssim 4$, as seen in Figs.
\ref{fig:mode_power_b3}  and  \ref{fig:Jz_evo}.   Furthermore,  the  CDI
external modes  can grow at the  higher latitude $|z|  \gtrsim 4$.  This
may be  because the strong $B_z$  shear in the radial  direction will be
relaxed towards  the jet  axis due to  the decreasing  external pressure
(expanding the poloidal fluxes to the radial direction).

\subsubsection{External Environment}
The  external  gas  itself  also  may  affect  the  stabilization.   The
inward-directed, thermal  pressure confinement and/or  the gravitational
contraction  of  the  external  unmagnetized gas,  which  surrounds  the
magnetic tower jets, may provide some stabilization at least against the
growth of  external CDI modes, although  the internal CDI  and PDI modes
will not  be affected  \citep[for the internal  PDI sausage  $m=0$ mode,
see][]{LF00}.  In  our results,  the  external  environment  may have  a
similar  stabilizing effect  on the  external CDI  modes to  the $B_{z}$
shear, as discussed in the previous section (\S \ref{sec:DS_RFP}), while
the internal PDI mode may be suppressed by the low internal pressure. In
the magnetic  tower jets, an  external gas plays  a crucial role  in the
stabilization of the CDI external  modes.  When an external gas pressure
profile does not decrease (constant),  which is examined in Paper I (see
Fig.  11 in Paper I for a continuous injection case), the magnetic tower
never expands radially and both  the thermal confinement by the external
gas  and the strong  $B_z$ shear  will be  kept throughout  the magnetic
tower ``body'' (no formation  of ``lobe'').  Therefore, the CDI internal
modes may occur, but the  CDI external modes should be surpressed. These
are consistent with the  CDI ``non-disrupting'' model \citep[]{A96, A00,
L00} as well.

\subsubsection{Rotation} 
We finally discuss the stabilization  by the rotation.  In our numerical
treatment,  we investigate almost  ``non-spinning'' magnetic  tower jet.
From  the theoretical point  of view,  any magnetically  driven outflows
(magneto-centrifugal  and/or  magnetic  pressure  driven)  can  have  an
azimuthal velocity component.  They  possess an angular momentum provided
by the accretion disk or Kerr black hole magnetosphere. In the classical
model of nonrelativistic MHD winds,  the centrifugal force plays a minor
role  in  the radial  force  equilibrium  far  from the  accretion  disk
\citep[]{BP82}.  However,  this is only the case  in the nonrelativistic
treatment.   In  the relativistic  MHD  regime,  the  situation will  be
modified;  semi-analytical  solutions  of  relativistic MHD  winds  with
self-similarity exhibit that  the relativistic inertia and electrostatic
forces play  an important  role in the  radial force equilibrium  at the
cylindrical   flow  structure,   extended   to  pc   or  larger   scales
\citep[]{VK03}.  Braided emission-line  profiles and rotational velocity
in  NGC 4258 indicate  a pure  helical motion  exists along  the tightly
wrapped strand even in kpc scales \citep[]{CWT92}.

The  stabilizing effect of  rotation against  the CDI  kink mode  in the
spinning nonrelativistic PFD/KFD jets has been investigated by numerical
simulations  \citep[]{NM04}.  Because of  the centrifugal  effect, these
spinning jets can be stabilized  against the $m=1$ mode beyond the point
predicted by  the K-S criterion.   The linear stability analysis  of the
relativistically   rotating   force-free   fields   in   Kerr   geometry
\citep[]{BZ77} has been performed  by \citet[]{T01}. They also show that
the field-line rotation has a  stabilizing effect against the kink modes
satisfying  the K-S  criterion.   In  these cases,  the  presence of  an
azimuthal velocity  component can  modify the radial  force equilibrium,
making the  jet robust against the  CDI kink mode.   Examinations of the
stability  property  in  a  rotating  magnetic  tower  jet  are  clearly
desirable.

\subsection{External CDI Kink Mode as a Possible Mechanism for Wiggling AGN Jets}
The  CDI  of current-carrying  jets  for  two  types of  return  current
distributions have been discussed: (1)  the return current is assumed to
flow  within the  jet itself  (the jets  are thermally  confined  by the
external medium) \citep[]{CPT89}, and  (2) the return current is assumed
to flow  around the  jet, such as  in a magnetized  cocoon \citep[]{B78,
B06}.  Most of the theoretical/numerical  works have been devoted to the
former case, in which the forward current $\bolJ^{\rm F}$ and the return
current $\bolJ^{\rm  R}$ are concentrated at the  interfaces between the
jet  and  ISM/IGM, which  is  taken  to  be ``unmagnetized''.   In  this
situation,  the  growth  of   external  CDI  modes  can  be  effectively
suppressed,  while  the  excitations  of  only internal  CDI  modes  are
allowed.   There  is   no  radial  displacement  at  the   edge  of  the
current-carrying column by the internal CDI modes, and therefore, it may
be difficult to explain the off-set axial distortion of observed jets by
these internal modes.  As described  in \S \ref{sec:INT}, this is one of
the reasons why much less attention has been paid to the CDI than to the
KHI  for the  stability problems  in astrophysical  jets to  explain the
disrupted jet exterior.

However,  in the  gravitationally stratified  atmosphere, the  effect of
thermal confinement on the edge  of the current-carrying column will get
gradually weaker as the magnetic tower grows.  As a result, a separation
of  current flowing paths  between $\bolJ^{\rm  F}$ and  $\bolJ^{\rm R}$
occurs, as seen  in Fig.  \ref{fig:Jz_evo}.  This implies  that the edge
of  $\bolJ^{\rm  F}$ {\em  becomes  a  free  boundary} against  the  CDI
external modes,  even when  it is embedded  deeply inside  the external,
unmagnetized  gas.   Furthermore,  the  region between  the  outside  of
$\bolJ^{\rm  F}$  and  the  inside  of $\bolJ^{\rm  R}$  becomes  nearly
force-free $\bolJ  \times \bolB \simeq 0$.  Therefore,  the CDI external
modes  can  grow  in a  dynamical  time  scale  to form  the  systematic
wiggles. This is  also true even in the spinning  PFD/KFD jets under the
large-scale magnetic field \citep[]{N01, NM04}.  On the other hand, even
if the gas pressure in the  external, unmagnetized gas is high enough to
suppress the growth of the CDI external kink mode, then the CDI internal
kink mode can  grow to form the internal  non-axisymmetric structure, as
we exhibited in Fig.   \ref{fig:Jz_evo}.  Therefore, we believe that the
external CDI  modes are  one of the  possible mechanisms to  explain the
disruption of current-carrying  MHD jets and it could  be applied to the
wiggling AGN jets.

\section{CONCLUSION}
Stability  properties  of magnetic  tower  jets  have  been examined  by
performing  3-D MHD  simulations. Our  numerical results  show  that the
magnetic tower  jets in the gravitationally  stratified atmosphere could
survive the current-driven kink instability beyond the Kruskal-Shafranov
criterion  even  in  the  nonlinear  regime.   One  of  the  potentially
stabilizing effects  on a traditionally  unstable magnetic field  may be
due to  the dynamical relaxation  of magnetic twists in  the propagating
magnetic tower jets.  This causes an increased threshold of the unstable
critical wavelength.

However,  the  magnetic  tower  jets  are eventually  distorted  by  the
non-axisymmetric  perturbations with a  few percent  of the  local sound
speed in  the system.  The current-driven kink  modes grow predominantly
on  time  scales  of the  order  of  the  local Alfv\'en  crossing  time
$\tau_{\rm A}$.  In the gravitationally stratified atmosphere, two types
of  kink modes appears:  the internal  and external  modes.  At  a large
distance away  from the  central region, the  external kink  mode grows,
while  only the  internal kink  mode exhibits  near the  central region.
Therefore, the exterior  of magnetic tower jets will  be deformed into a
large-scale wiggled structure by the external kink mode.

Possible mechanisms  of stabilizing/destabilizing in  the magnetic tower
jets    under   the    gravitationally   stratified    atmospheres   are
discussed. None  of the  growing surface/body modes  of Kelvin-Helmholtz
instability are identified due to the small velocity shear and slow bulk
speed  (compared with  the  local Alfv\'en  speed). The  pressure-driven
instabilities  are also  inhibited  due  to the  low  plasma $\beta$  ($
\lesssim 0.1$) inside the  magnetic towers. The ``reversed field pinch''
field configuration in the towers  and the external unmagnetized gas may
provide some stabilization effects on the current-driven external modes.

Non-axisymmetric current-driven  instabilities ($m \ge  1$) are absolute
instabilities, i.e., they grow but do not propagate in the jet co-moving
(rest)  frame   \citep[]{A00,  NM04}.   If   the  dissipation  timescale
$\tau_{\rm d}$ of  the current associated with jets  is much longer than
the  jet  dynamical  timescale  $\tau_{j}  \lesssim  \tau_{\rm  A}  \sim
\tau_{\rm CDI} \ll \tau_{\rm d}$,  the patterns created by the CDI modes
will persist  for some  time as  the flow advances.   And the  bulk flow
itself will appear to travel in a true 3-D helical pattern as it follows
the  magnetic backbone  of  the helix,  as  observationally expected  in
several radio sources as 3C 345 \citep[]{Z95}, 3C 120 \citep[]{G01}, and
3C 449 \citep[]{F99}.

\acknowledgments  Helpful  discussions   with  J.  Finn  are  gratefully
acknowledged.   This work  was carried  out  under the  auspices of  the
National  Nuclear  Security Administration  of  the  U.S. Department  of
Energy   at  Los   Alamos   National  Laboratory   under  Contract   No.
DE-AC52-06NA25396.  It was supported by the Laboratory Directed Research
and Development Program at LANL and by IGPP at LANL.

\end{document}